\documentclass[lettersize,journal]{IEEEtran}
\usepackage{amsmath,amsfonts}
\usepackage{algorithmic}
\usepackage{algorithm}
\usepackage{array}
\usepackage{textcomp}
\usepackage{stfloats}
\usepackage{url}
\usepackage{verbatim}
\usepackage{graphicx}
\usepackage{cite}
\usepackage{siunitx}
\usepackage{multirow}
\usepackage{booktabs}
\usepackage{amssymb}
\usepackage{xcolor}
\usepackage{makecell}
\usepackage{subcaption}
\usepackage[colorlinks,linkcolor=black,urlcolor=black,citecolor=black]{hyperref}

\begin{document}
\title{WavCaps: A ChatGPT-Assisted Weakly-Labelled Audio Captioning Dataset for Audio-Language Multimodal Research}

\author{Xinhao Mei,
    Chutong Meng,
    Haohe Liu,
    Qiuqiang Kong,\\
    Tom Ko,
    Chengqi Zhao,
    Mark D. Plumbley,
    Yuexian Zou,
    Wenwu Wang
\thanks{X. Mei, H. Liu, M. D. Plumbley, and W. Wang are with the
Centre for Vision, Speech, and Signal Processing, University of Surrey, Guildford, GU2 7XH, U.K. (E-mail: [x.mei, haohe.liu, m.plumbley, w.wang]@surrey.ac.uk)}
\thanks{C. Meng is with Johns Hopkins University, U.S.A. (E-mail: cmeng9@jhu.edu)}
\thanks{Q. Kong is with The Chinese University of Hong Kong, Hong Kong, China. (E-mail: qqkong@ee.cuhk.edu.hk)}
\thanks{T. Ko, and C. Zhao are with ByteDance, China. (E-mail: 
 [tom.ko, zhaochengqi.d]@bytedance.com)}
\thanks{Y. Zou is with the School of Electronic and Computer Engineering, Peking University, Shenzhen Graduate School, Shenzhen, 518055, China. (E-mail: zouyx@pku.edu.cn)}
}



\maketitle

\begin{abstract}
The advancement of audio-language (AL) multimodal learning tasks has been significant in recent years, yet the limited size of existing audio-language datasets poses challenges for researchers due to the costly and time-consuming collection process. To address this data scarcity issue, we introduce \textit{WavCaps}, the first large-scale weakly-labelled audio captioning dataset, comprising approximately 400k audio clips with paired captions. We sourced audio clips and their raw descriptions from web sources and a sound event detection dataset. However, the online-harvested raw descriptions are highly noisy and unsuitable for direct use in tasks such as automated audio captioning. To overcome this issue, we propose a three-stage processing pipeline for filtering noisy data and generating high-quality captions, where ChatGPT, a large language model, is leveraged to filter and transform raw descriptions automatically. We conduct a comprehensive analysis of the characteristics of WavCaps dataset and evaluate it on multiple downstream audio-language multimodal learning tasks. The systems trained on WavCaps outperform previous state-of-the-art (SOTA) models by a significant margin. 
Our aspiration is for the WavCaps dataset we have proposed to facilitate research in audio-language multimodal learning and demonstrate the potential of utilizing large language models (LLMs) to enhance academic research.
Our dataset and codes are available at \url{https://github.com/XinhaoMei/WavCaps}.

\end{abstract}


\section{Introduction}
\label{sec:intro}
\IEEEPARstart{O}{ver} the past decade, the field of machine listening has achieved notable progress, with the aid of deep learning techniques and the availability of audio event datasets \cite{audioset, piczak2015esc50, fsd50k}. This has resulted in the development of algorithms that can detect and identify sound events and acoustic scenes \cite{kong2019weakly, mesaros2021sed, Xu2017unsupervised, barchiesi2015acoustic}.
More recently, there has been a surge of interest in establishing a more profound comprehension of audio content by connecting audio and language. A number of audio-language (AL) multimodal learning tasks have been introduced, such as text-to-audio retrieval \cite{xie2022atretrieval, koepke2022audioretrieval, mei2022metric}, automated audio captioning \cite{mei2022ac_review, Gontier2021ac_bart, koizumi2020keywords}, audio question answering \cite{lipping2022clotho-aqa, fayek2020aqatemp}, text-based sound generation \cite{kreuk2022audiogen, liu2023audioldm, yang2022diffsound, agostinelli2023musiclm}, and text-based sound separation \cite{liu2022separate}. Research on AL takes a stride in the direction of empowering machines to comprehend audio signals at a human-like level. 

While AL research is relatively young, vision-language (VL) multimodal learning \cite{uppal2022multimodal}, the counterpart of AL multimodal learning, has been studied for decades and has contributed to the success of many vision-language applications, such as cross-modal search \cite{qi2020imagebert}, image and video captioning \cite{zhou2020unifiedVLP, li2020oscar}, text-to-image generation \cite{ramesh2022hierarchicalt2i, Rombach_2022diffusion_t2i}, and visual question answering \cite{antol2015vqa, yu2019deep_vqa}. Two main factors have contributed to the significant progress in VL tasks. First, advances in model architectures, especially self-attention-based Transformer models \cite{vaswani2017attention}, have shown superior performance compared to convolutional neural networks (CNNs) \cite{lecun2015cnn} and recurrent neural networks (RNNs) \cite{rumelhart1986rnn}, both in computer vision and natural language processing tasks. Second, large-scale VL multimodal pretraining enables models to learn transferable and robust multimodal representations that benefit VL downstream tasks \cite{li2021albef, li2022blip, li2020oscar}. Pre-training on large-scale (even weakly-labelled) datasets and fine-tuning on specific datasets for downstream tasks has been a prevalent methodology in VL multimodal learning tasks \cite{jia2021align, radford2021clip}.  

Inspired by the progress made in the domains of vision and language, Transformer-based models and training strategies have been successfully adapted in modeling audio signals \cite{gong2021ast}. This has led to substantial improvements in the recognition and detection of sound events. However, the scale of audio-related datasets is still limited, which severely hinders the research in AL multimodal learning. Till now, the largest audio event dataset, AudioSet \cite{audioset}, has about 2M audio clips, and the largest audio captioning dataset, AudioCaps \cite{kim2019audiocaps}, contains only about 50k audio clips. Both of them are orders of magnitude smaller than their counterparts in the vision domain such as ImageNet \cite{imagenet_cvpr09}, and MS COCO \cite{chen2015mscoco}. The reasons for this are twofold: First, the research within the audio community has received less attention, as compared to the vision community. Second, the collection of audio datasets is a more laborious, costly, and time-consuming process compared to that of visual datasets.

To alleviate the data scarcity problem and advance AL research, we gathered audio clips and their corresponding raw descriptions from multiple sources, drawing inspiration from the collection process of the Conceptual Captions datasets \cite{sharma2018cc3m, changpinyo2021cc12m}. Our collection encompasses data from three web platforms—FreeSound\footnote{\url{https://freesound.org/}}, BBC Sound Effects\footnote{\url{https://sound-effects.bbcrewind.co.uk/}}, and SoundBible\footnote{\url{https://soundbible.com/}}—as well as from an audio tagging dataset, AudioSet. These comprehensive and varied sources are instrumental to enriching the depth and breadth of our dataset. Nevertheless, the harvested raw descriptions may range from complete sentences to fragmented phrases, lists of keywords or tags, and exhibit a high degree of noise. For instance, some descriptions do not represent the audio content at all, and certain descriptions contain extraneous information unrelated to the audio, such as recording devices, time, and locations. As a result, these unrefined raw descriptions hinder the learning of robust audio-language representations and are ill-suited for direct application in tasks like automated audio captioning. To process harvested data, Conceptual Captions 3M (CC3M) \cite{sharma2018cc3m} applied a series of complex rules to filter and transform candidate image-caption pairs and it only keeps around 0.2\% of originally harvested data. When compared to image-description pairs, the limited number of audio-description pairs available on the web makes it impractical to accept such a high rate of discarding.

To overcome this problem, we propose a three-stage processing pipeline for filtering noisy data and generating high-quality captions. In a pioneering move, we utilize ChatGPT\footnote{\url{https://openai.com/blog/chatgpt/}}, a robust large language model to automate the process, ensuring both efficiency and effectiveness. Initially, a pre-filtering stage is implemented to exclude irrelevant data based on text frequency. Subsequently, ChatGPT is employed to further process the obtained raw descriptions. This involves content-based filtering and transforming the raw descriptions into sentences resembling captions. Lastly, we refine undesirable outputs from the second stage in a post-processing stage. Ultimately, we present the first large-scale, weakly-labelled audio captioning dataset, WavCaps, which contains about \num{400}k audio clips with paired captions. 

We classify WavCaps as a weakly-labeled dataset for the following reasons. The term ``weakly-labeled dataset" is derived from the domain of weak supervision or weakly supervised learning \cite{zhou2017weakly}. This area of study addresses situations where the supervision signal provided for model training is imperfect \cite{zhou2017weakly, Li2021weakly},  covering the scenarios such as incomplete supervision, where only a portion of the training data has labels; inexact supervision, where the labels used are not as granular as the task demands; and inaccurate supervision, where the labels are tainted with noise or errors. In WavCaps, captions are derived by filtering and transforming harvested raw descriptions using ChatGPT. They may be incomplete if specific sound events are absent from the raw descriptions or incorrect if the descriptions themselves are erroneous, thus exemplifying the characteristics of inexact and inaccurate supervision in weakly supervised learning.

In comparison to existing audio captioning datasets \cite{kim2019audiocaps, drossos2020clotho, Martin2021databias}, WavCaps is not only an order of magnitude larger, but also encompasses a wider range of content. We conducted experiments on multiple audio-language multimodal learning tasks to evaluate the impact of the proposed WavCaps dataset and achieved new state-of-the-art results on most tasks, surpassing previous benchmarks by significant margins.


In summary, our work offers three main contributions: (1) the introduction of a large-scale, weakly-labeled audio captioning dataset, WavCaps, for audio-language multimodal learning tasks; (2) the use of ChatGPT to automatically filter and rewrite harvested raw descriptions into caption-like sentences, showcasing its powerful data augmentation capabilities; and (3) extensive experiments conducted on downstream tasks, demonstrating the effectiveness of our proposed WavCaps dataset. Our expectation is that WavCaps will aid in advancing research in audio-language multimodal learning and also serve as a demonstration of how ChatGPT can be utilized to enrich academic research. 

The remainder of this paper is organized as follows. Section~\ref{sec:related_works} introduces related works in vision-language and audio-language areas. Details of the dataset collection and processing steps are described in Section~\ref{sec:wavcaps_dataset}. Section~\ref{sec:exp} introduces experiments on audio-language multimodal tasks and present the results and analysis. Finally, we conclude this work in Section\ref{sec:conclu}.

\begin{figure*}[ht]
  \centering
  \includegraphics[width=0.85\textwidth]{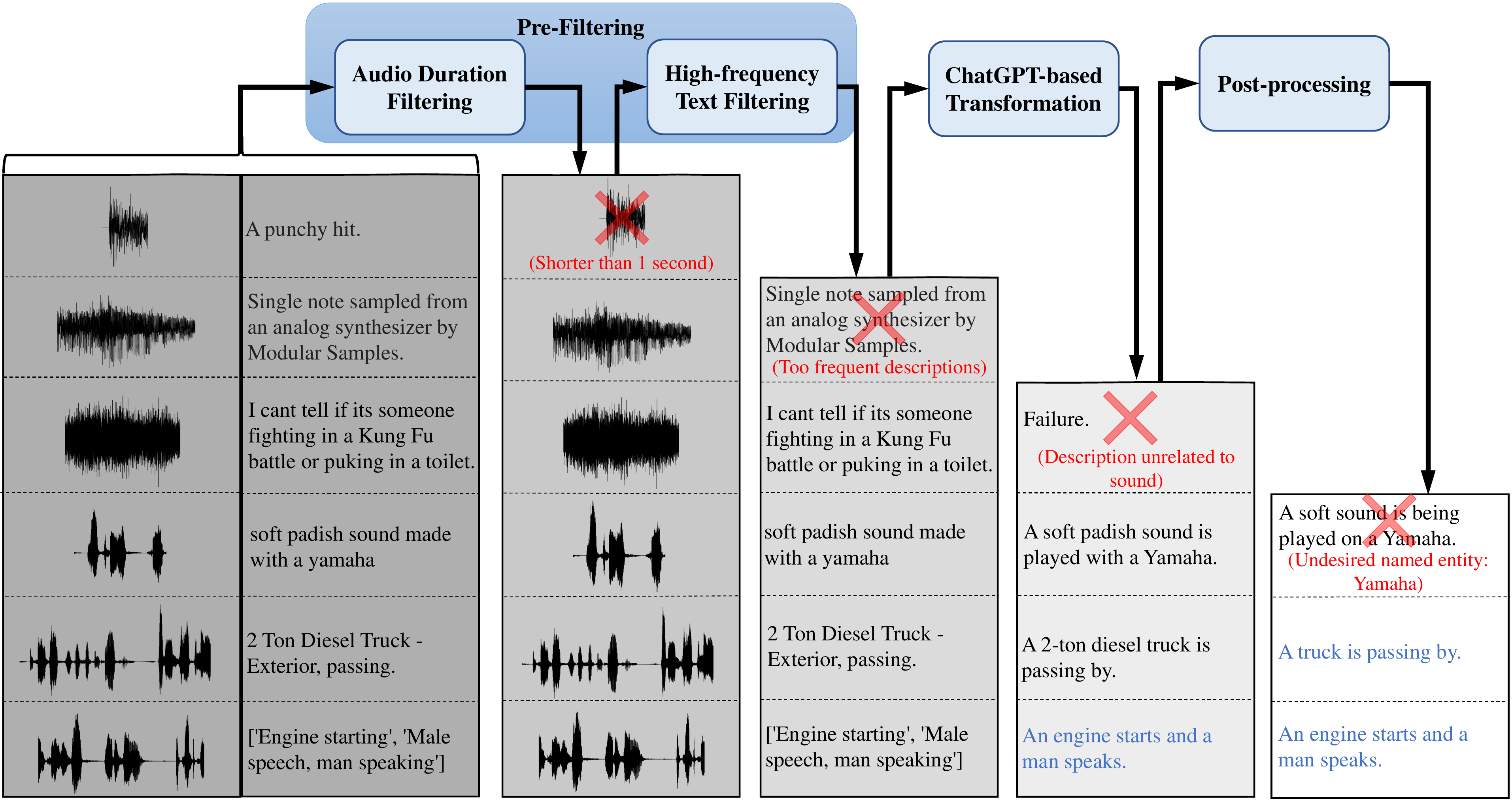}
  \caption{Overview of the three-stage data processing pipeline.}
  \label{fig:pipeline}
\end{figure*}

\section{Related Works}
\label{sec:related_works}
\subsection{Vision-Language Datasets}
In recent years, significant progress has been made in VL learning thanks to the release of large-scale VL datasets. The images or videos in these datasets are typically sourced from online platforms. Based on the annotation methods employed, VL datasets can be classified into two categories: automatically-annotated and human-annotated. Automatically-annotated datasets, such as CC3M \cite{sharma2018cc3m}, Conceptual Captions 12M (CC12M) \cite{changpinyo2021cc12m}, and ALIGN \cite{jia2021align}, have millions of image-caption pairs, where image-description pairs are first harvested from the web and processed automatically according to a series of predefined rules, without human intervention. These large-scale datasets are usually used as pre-training datasets to learn multimodal representations for downstream tasks. In contrast, human-annotated datasets, such as COCO Captions \cite{chen2015mscoco}, NOCaps \cite{agrawal2019nocaps}, and Flickr\cite{young2014flickr}, employ humans to annotate images or videos. The annotation process is expensive and time-consuming. Therefore, human-annotated datasets are limited in size and are generally used as fine-tuning datasets for performance evaluation. In summary, automatically-annotated datasets are larger but often noisy, while human-annotated datasets are smaller but of better quality. Pre-training on large-scale (even noisy) datasets and fine-tuning on small task-specific datasets has been a prevalent methodology in VL multimodal learning \cite{jia2021align, li2022blip, li2021albef}.

Similar to automatically-annotated VL datasets \cite{sharma2018cc3m, changpinyo2021cc12m, jia2021align}, we also harvest audio-description pairs from the web, and design a pipeline to process data automatically. However, taking into account the differences in quantity and data characteristics, we did not use complex pre-defined rules as in CC3M \cite{sharma2018cc3m} to process the harvested data, but  employed ChatGPT to filter and rewrite raw descriptions into sentence-like captions.

\subsection{Audio-Language Datasets}
Compared with flourishing research on VL multimodal learning, research on audio-language multimodal learning is limited due to the lack of AL datasets. Almost all AL tasks, such as automated audio captioning \cite{mei2022ac_review, Mei2021ACT, liu2022leveraging}, language-based audio retrieval \cite{koepke2022audioretrieval, mei2022metric}, text-to-audio generation \cite{kreuk2022audiogen, liu2023audioldm} and language-queried sound separation \cite{liu2022separate}, rely on human-annotated audio captioning datasets, AudioCaps \cite{kim2019audiocaps} and Clotho \cite{drossos2020clotho, lipping2019crowdsourcing}. Collecting human-annotated datasets is expensive and time-consuming, necessitating meticulously planned collection processes, such as determining the information to be made available to human annotators (e.g., visual aids), and careful post-processing to ensure their quality and accuracy. AudioCaps is the largest human-annotated audio captioning dataset, containing about 50k audio clips sourced from AudioSet \cite{audioset}, the largest audio event dataset. The training set of AudioCaps has one human-annotated caption per audio clip, while the validation and test sets have five human-annotated captions per audio clip. Clotho contains around 6k audio clips sourced from the FreeSound platform, each audio clip has five human-annotated captions. Although other human-annotated audio captioning datasets have been proposed, such as MACS \cite{Martin2021databias} and AudioCaption \cite{wu2019audiocaption_hospital}, they are still limited in size and not widely used due to their inferior quality compared to AudioCaps and Clotho. The methodologies and post-processing procedures of these human-annotated datasets vary significantly. For more detailed information about these datasets, we recommend readers to refer to the respective papers. The insufficient size of these human-annotated audio-language datasets poses a significant obstacle to AL research.

Other audio researchers have also tried to harvest audio clips and their descriptions from the web. Koepke et al. \cite{koepke2022audioretrieval} introduced SoundDescs dataset, which is sourced from BBC Sound Effects archive. Soham et al. \cite{deshmukh2022wavtext5k} crawled 5k audio clips and their descriptions from two sources on the web and released WavText5K for language-based audio retrieval task. Wu et al. \cite{laionclap2023} introduced LAION-Audio-630K, the largest audio-language dataset to date, by crawling audio and descriptions from multiple sources, where most of the audio clips come from Freesound. In addition, Wu et al. investigated to employ T5 \cite{colin2020t5}, a pre-trained large language model, to generate captions based on ground truth tags of audio clips in AudioSet. Much like LAION-Audio-630K, the majority of the data included in our WavCaps dataset also originates from Freesound. There are two main differences between our work and other online-harvested AL datasets. First, raw descriptions harvested from the web are very noisy. Existing harvested AL datasets did not filter or process these raw descriptions. Instead, we utilized ChatGPT to sift through and rephrase the unprocessed descriptions into sentences that resemble captions, making our WavCaps dataset suitable for all kinds of audio-language tasks including automated audio captioning. Second, classes in AudioSet are very unbalanced and most of the audio clips only have a single label \cite{kong2020panns}, therefore, using T5 to transform tags to captions makes the training data unbalanced and noisy. We propose to create captions for the subset of AudioSet \cite{hershey2021audioset_sed} that contains precise timing information for sound events, as labeled by humans.

Since the release of our WavCaps dataset in March 2023, our methodology has inspired a variety of subsequent research works and WavCaps has been used to train diverse audio-language models. These include the creation of new datasets \cite{gong2023listen, sun2023large, doh2023lp}, the development of text-to-audio generation models \cite{yang2023uniaudio, liu2023audioldm2, ghosal2023tango}, and the advancement of large audio language models \cite{tang2023salmonn, zhao2023chatbridge, liang2023apt}. This breadth of research demonstrates the significant impact and ongoing relevance of our work in the evolving field of audio-language multimodal learning.

\section{WavCaps Dataset}
\label{sec:wavcaps_dataset}
In this section, we introduce the collection and processing steps of the WavCaps dataset. Firstly, we provide an introduction to the data sources and their respective characteristics. We then describe our proposed three-stage processing pipeline, including pre-filtering, ChatGPT-based transformation, and post-processing. Finally, we present a detailed analysis of the WavCaps dataset. Fig.~\ref{fig:pipeline} shows the overview of the three-stage data processing pipeline.

\begin{table*}[!t]
\caption{Example prompts to ChatGPT for FreeSound and AudioSet strongly-labelled subset. Transformation examples are are ignored. `SL' refers to `strongly-labelled'.}
\centering
\resizebox{\textwidth}{!}{
\begin{tabular}{c|c}
\hline
\textbf{Data sources}  & \textbf{Prompts} \\
\hline
\makecell[{{}}]{FreeSound \\  \\BBC Sound Effects\\ \\ SoundBible}  &  \makecell[{{p{0.8\textwidth}}}]{I will give you a list of descriptions of sounds. Process each individually. Extract the type of the sound and generate an audio caption describing the sound events. The audio caption should be less than 20 words. Delete the author of the sound. Delete locations, city names, country names. Delete the time. Delete device names. Delete the proper noun modifiers, number modifiers, and unit modifiers. Summarize each output into one sentence. Replace all named entities with their hypernyms. Replace people names with ``someone''.\\
Do not write introductions or explanations. Only describe the sound events and do not use ``heard'', ``recorded''. Start each output sentence with its index. Make sure you are using grammatical subject-verb-object sentences. Output ``Failure.'' if the description is not related to sound.} \\
\hline
AudioSet SL & \makecell[{{p{0.8\textwidth}}}]{I will give you a number of lists containing sound events occurred sequentially in time. Process each individually. Write an one-sentence audio caption to describe these sounds.\\
Make sure you are using grammatical subject-verb-object sentences. Directly describe the sounds and avoid using the word ``heard''. The caption should be less than 20 words.} \\
\hline
\end{tabular}
}
\label{tab:prompts}
\end{table*}

\subsection{Data Sources}

\noindent \textbf{FreeSound} \cite{font2013freesound} is an online collaborative sound sharing site started in \num{2005}. The initial goal of FreeSound is to give support to sound researchers and sound artists who usually have trouble in finding royalty-free sound samples. After more than \num{10} years of development, there are more than \num{560000} audio clips\footnote{Including data up to the November 30th 2022.} uploaded by registered users, and these audio clips cover diverse contents such as music, environmental sounds, synthesized sound effects and even noises. When uploading an audio clip, each user is asked to give a short description as well as annotations of other attributes such as tags of the sound events about the uploaded audio clip. Ideally, we would like to use user-uploaded descriptions directly as audio captions. However, these raw descriptions are extremely noisy. For example, some are not related to the audio content (such as the recording device or personal feeling), some are not sentences but only nouns or verbs, and some are too detailed and include many specific information such as place names and people names. In addition, there are a large number of repetitive descriptions uploaded by the same user when they upload many audio clips at the same time. Due to these distinct attributes, processing descriptions sourced from FreeSound can prove to be extremely challenging.

\noindent \textbf{BBC Sound Effects} contains over 33k audio clips recorded around the world over the past 100 years. These audio clips contain extensive contents from the BBC Radiophonic workshop, the Blitz in London, BBC TV and Radio productions, and the BBC Natural History Unit archive. The raw descriptions in BBC Sound Effects begin with nouns describing the sounding objects or acoustic scenes of the sound content, followed by detailed descriptions. However, most of the detailed descriptions are also not complete sentences, and some include specific information such as recording equipment, time and places. Although these raw descriptions are noisy and cannot be directly used as audio captions, these raw descriptions all describe the content of the audio clips.

\begin{table*}[!t]
\caption{Examples of raw descriptions and ChatGPT's outputs.}
\centering
\resizebox{\textwidth}{!}{
\begin{tabular}{c|cc}
\hline
\textbf{Data sources}  & \textbf{Raw descriptions} &\textbf{ChatGPT's output} \\
\hline
 \multirow{2}{*}{FreeSound}  & \makecell[{{p{0.5\textwidth}}}] { This sound is of a book falling down the staircase in the Stanford University library west stacks.} & A book is falling down a staircase. \\
& \makecell[{{p{0.5\textwidth}}}] {After 2h and 15min. Excerpt.} & Failure. \\
\hline
\multirow{2}{*}{BBC Sound Effects}  & \makecell[{{p{0.5\textwidth}}}] {Timber \& Wood - Rip saw, carpenters' workshop.} & Someone is using a rip saw in a carpenter's workshop. \\
& \makecell[{{p{0.5\textwidth}}}] { Motor Car: Ford Ecsort 1300 (Automatic) - Ford Escort 1300, exterior, passing with horn. (1300cc engine, automatic transmission.)} & A car is passing with its horn. \\
\hline
\multirow{2}{*}{SoundBible}  & \makecell[{{p{0.5\textwidth}}}] {Tasmanian Devil growling screaming hissing. Warning sounds from a Tasmanian Devil in Zoo.} & An animal is growling, screaming, and hissing. \\
& \makecell[{{p{0.5\textwidth}}}] {Large Tibetan Bells ringing in a temple. Could also use for Monastery or Monks.} & Bells are ringing. \\
\hline
\multirow{2}{*}{AudioSet SL}  & \makecell[{{p{0.5\textwidth}}}] {[`Accelerating, revving, vroom', `Race car, auto racing']} & A race car is accelerating and revving. \\
& \makecell[{{p{0.5\textwidth}}}] {[`Female speech, woman speaking', `Whoosh, swoosh, swish']} & A woman is speaking while something whooshes. \\
\hline
\end{tabular}
}
\label{tab:examples}
\end{table*}

\noindent \textbf{SoundBible} is a website for sharing free and royalty free sound effects and audio clips. We harvested about \num{1500} royalty free sound effects with their raw descriptions from SoundBible, and these audio clips also cover a wide range of contents. Most of the raw descriptions in SoundBible are well-written sentences describing the audio content, but some of them still contain redundant information unrelated to the audio content. As a result, these raw descriptions remain unsuitable for direct use as audio captions.

\noindent \textbf{AudioSet Strongly-Labelled Subset} \cite{hershey2021audioset_sed} is a sound event detection dataset that is included to increase the size of our proposed WavCaps dataset. The original AudioSet dataset \cite{audioset} contains about 2M audio clips with unbalanced, weakly-labelled clip-level tags, and most of these audio clips only has one single tag. To investigate whether accuracy of the classifier trained on AudioSet is impaired by the weak labels, Shawn et al. \cite{hershey2021audioset_sed} have also made available a strongly-labelled subset of approximately 100k audio clips from the AudioSet dataset. This subset has been annotated by humans and includes precise timing information for the sound events that occur within each clip. With these strongly-labelled temporal information, template-based methods or large language models can be employed to generate captions for audio clips in AudioSet strongly-labelled subset.

\subsection{Data Processing}
\label{ssec: data_char}
Online-harvested raw descriptions are very noisy and thus cannot be directly used as captions. To address this issue, image captioning datasets, such as CC3M, and CC12M, have applied a series of complex filtering and transformation steps to process raw image-description pairs, including image-based filtering, text-based filtering, image and text-based filtering and text transformation. 
The processing steps are stringent, resulting in a significant discard rate that we cannot afford. In order to retain as much data as possible, we propose a three-stage processing pipeline by simplifying the filtering and transformation steps according to the characteristics of the harvested audio-description pairs, as shown in Figure~\ref{fig:pipeline}. 

\noindent \textbf{Pre-Filtering.}
Initially, we apply minimal pre-filtering to remove undesirable data, such as descriptions that do not pertain to audio content. The pre-filtering consists of audio-duration filtering and high-frequency text filtering. For audio-duration filtering, audio clips with a duration of less than one second are removed. This is mainly because short audio clips might not contain enough meaningful content and require extremely long padding during training. We limit our high-frequency text filtering solely to data obtained from FreeSound. This is because the raw descriptions from BBC Sound Effects and SoundBible are generally considered to be consistent with their corresponding audio content, and the sound event labels in the AudioSet strongly-labelled subset are manually annotated. Regarding the FreeSound data, we have noticed that high-frequency descriptions that are shared across multiple audio clips are typically uploaded by the same user who uploads multiple clips simultaneously. These descriptions are more likely to be unrelated to the audio content. Therefore, we apply high-frequency text filtering to exclude descriptions that are shared by more than \num{5} audio clips. These two filtering steps have removed about \num{265}k data samples from FreeSound.

\begin{table*}[!t]
\caption{Statistics of harvested raw data and WavCaps dataset.}
\centering
\begin{tabular}{c|ccc|ccc}
\hline
\multirow{2}{*}{\textbf{Data sources}} & \multicolumn{3}{c}{\textbf{Before processing}} & \multicolumn{3}{c}{\textbf{After processing}}\\
    \cline{2-7}
    & \textbf{num. of audio} &\textbf{avg. audio duration (s)} & \textbf{avg. text length} & \textbf{num. of audio} &\textbf{avg. audio duration (s)} & \textbf{avg. text length}\\
\hline
FreeSound & 567078 & 56.87 & 17.74 & 262300 & 85.98 & 6.77 \\
BBC Sound Effects & 33064 & 115.75 & 15.91 & 31201 & 115.04 & 9.67 \\
SoundBible & 1576 & 11.20 & 17.90 & 1232 & 13.12 & 5.87 \\
AudioSet SL subset & 108317 & 10.00 & - & 108317 & 10.00 & 9.79 \\
\hline
WavCaps & - & - & - & 403050 & 67.59 & 7.80 \\
\hline
\end{tabular}
\label{tab:process_statistics}
\end{table*}

\noindent \textbf{ChatGPT-based Transformation.}
In light of the previously discussed characteristics of the raw data, we propose three fundamental principles that we believe are essential for effectively converting raw metadata into audio captions:
\begin{itemize}
\item Use a single-sentence, accurate description of the audio's content using concise syntax that follows the grammatical subject-verb-object structure;
\item Avoid the use of named entities such as individual names, locations, and recording devices, which are not related to the audio content;
\item Omit any irrelevant subjective information, like personal emotions or opinions, that is unrelated to the sound.
\end{itemize}
Given the varied nature of raw descriptions, it is challenging to apply conventional rule-based natural language processing methods as in CC3M \cite{sharma2018cc3m} or CC12M \cite{changpinyo2021cc12m}. Attempting to do so may result in a high percentage of metadata and audio clips being discarded, comparable to that observed in CC3M. 

To tackle the challenge of converting raw descriptions into captions, we propose using ChatGPT, a powerful conversational large language model trained by OpenAI\footnote{\url{https://openai.com/}} to perform this task automatically.
Unlike traditional large language models such as T5 \cite{colin2020t5} used in LAION-Audio-630K \cite{laionclap2023}, ChatGPT has been shown to excel at generating human-like responses to natural language prompts, and has garnered widespread attention for its powerful understanding, reasoning, and dialogue abilities. By designing prompts that account for the characteristics of  different data sources, ChatGPT can effectively filter out sound-unrelated information and rewrite raw descriptions in to audio caption-like sentences that meet the requirements we proposed in prompts. This approach has the potential to significantly reduce the discard rate of raw descriptions and improve the quality of converted captions. Prompts we used are shown in Table~\ref{tab:prompts}. In order to make use of ChatGPT's in-context learning ability, several transformation examples are also included in the prompts and they are different for each data source (ignored in Table~\ref{tab:prompts}). These examples can significantly improve the caption quality. It is worth noting that other advanced conversational LLMs emerged recently, such as GPT-4 and LLaMA3 \footnote{\url{https://llama.meta.com/}}, could also be used. In terms of our empirical tests, however, the captions generated by these models retain similar semantic information, despite their variations in the choice of the words.

Table~\ref{tab:examples} presents examples of the raw descriptions and final processed captions. It can be observed that ChatGPT can transform fragmented descriptions (e.g., nouns and phrases) into sentences, remove redundant information that is either too specific or not related to sound, and condense lengthy sentences into more concise captions. This capability enhances the usability of the data and can be considered a form of data augmentation, enriching the dataset with more coherent audio captions. 
However, ChatGPT also serves as a critical quality control measure, ensuring the relevance and accuracy of the dataset, due to its ability to output a ``Failure'' message for the descriptions that are not related to audio content. This ability, distinct from data augmentation, focuses on selectively refining the dataset by removing irrelevant or misleading entries, thereby enhancing the overall data quality and utility.

While ChatGPT has shown promising results in converting raw descriptions into captions, in some cases, it may still fail to follow the prompts to produce captions that meet our requirements. For instance, ChatGPT may encounter difficulty in removing numbers, names of individuals, and geographic locations from raw descriptions. Moreover, a small percentage of descriptions that have no relation with the audio content may not be filtered out. 

\begin{figure*}[!htbp]
    \centering
    \begin{subfigure}[b]{0.32\textwidth}
        \includegraphics[width=\textwidth]{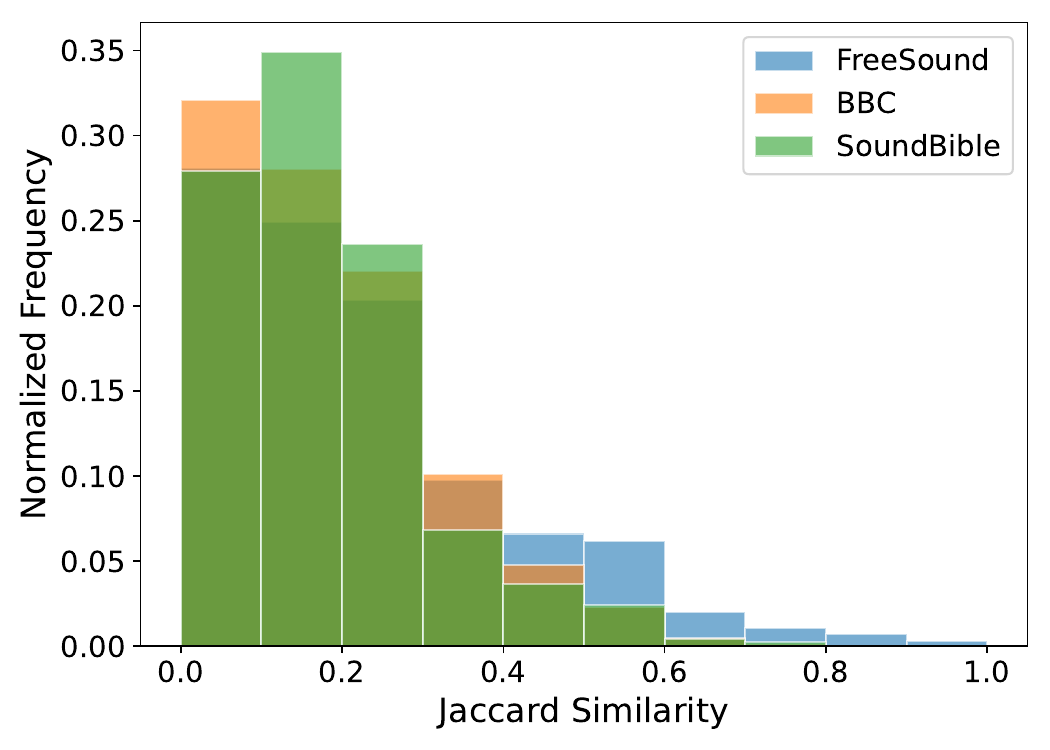}
        \caption{ }
        \label{fig:jaccard_similarities}
    \end{subfigure}
    \hfill
    \begin{subfigure}[b]{0.27\textwidth}
        \includegraphics[width=\textwidth]{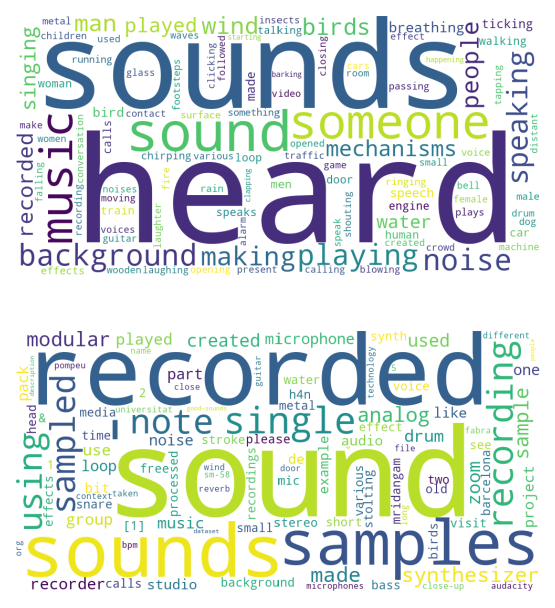}
        \caption{ }
        \label{fig:word_clouds}
    \end{subfigure}
    \hfill
    \begin{subfigure}[b]{0.3\textwidth}
        \includegraphics[width=\textwidth]{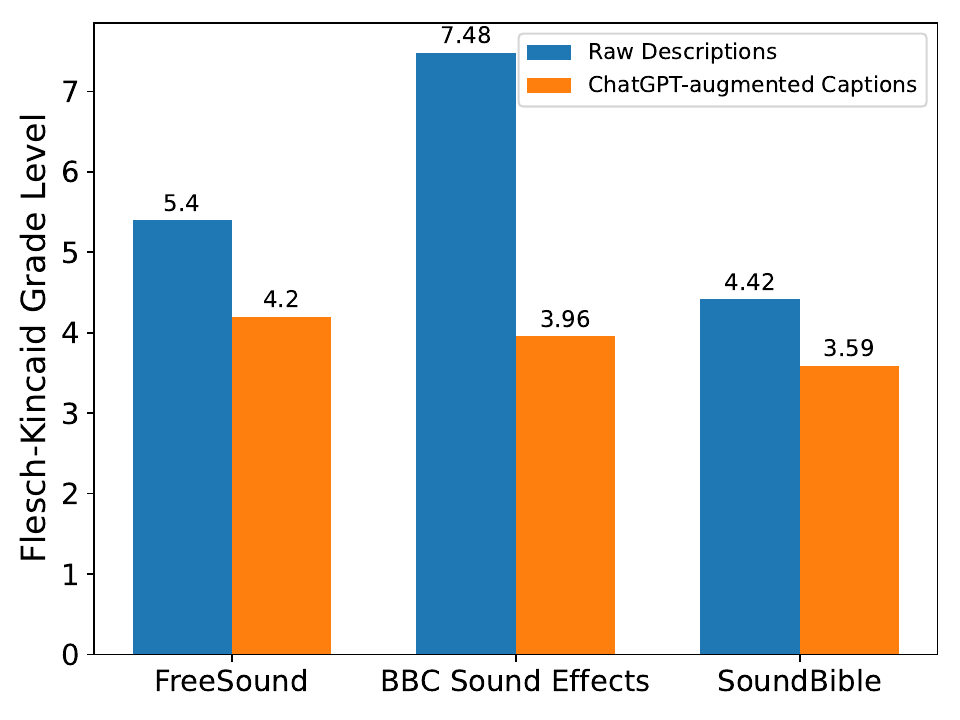}
        \caption{ }
        \label{fig:flesch_kincaid}
    \end{subfigure}

    \caption{Comparative visualization of raw descriptions between ChatGPT-augmented captions, Figure \ref{fig:jaccard_similarities} shows the Jaccard similarity between raw descriptions and ChatGPT-augmented captions, Figure \ref{fig:word_clouds} presents word clouds of top 100 words in the WavCaps dataset (top) and the entire harvested raw descriptions (bottom), stop words are ignored, and Figure \ref{fig:flesch_kincaid} displays the readability of raw descriptions and ChatGPT-augmented captions as indicated by Flesch-Kincaid grade levels.}
    \label{fig:comparision_combined_figures}
\end{figure*}



\noindent \textbf{Post-Processing.} To address the aforementioned cases of incorrect processing, we implement post-processing steps to refine the captions generated by the system. To identify erroneous outputs that still include numbers, geographic locations, and individuals' names, we utilize a pre-trained named entity recognition model from spaCy \footnote{https://spacy.io/models/en\#en\_core\_web\_sm}. Any captions with named entity information are then processed again by ChatGPT to remove these details using the same prompts but different examples. In most cases, this second round of processing successfully removes the named entity information. However, if post-processing still results in captions with named entity information, we discard these captions to ensure the quality of the final dataset. Finally, we exclude captions that are too brief by setting a minimum length threshold of three words. This ensures that the captions are informative and descriptive enough to convey a meaningful description of the audio content.

\begin{table}[t]
\caption{Comparative overview of main audio-language datasets between our proposed WavCaps dataset.}
\centering
\resizebox{\linewidth}{!}{
\begin{tabular}{c|ccc}
\hline
\textbf{Dataset}  & \textbf{Num. audios} &\textbf{Duration (h)} & \textbf{Text source} \\
\hline
AudioCaps \cite{kim2019audiocaps} & 52904 & 144.94 & Human \\
Clotho \cite{drossos2020clotho} & 5929 & 37.00 & Human \\
MACS \cite{Martin2021databias} & 3537 & 9.83 & Human \\
WavText5K \cite{deshmukh2022wavtext5k} & 4072 & 23.20 & Online raw-data \\
SoundDescs \cite{koepke2022audioretrieval} & 32979 & 1060.4 & Online raw-data \\
LAION-Audio-630K \cite{laionclap2023} & 633526 & 4325.39 & Online raw-data \\
\hline
WavCaps & 403050 & 7567.92 & ChatGPT \\
\hline
\end{tabular}
}
\label{tab:dataset_statistics}
\end{table}

\begin{figure}[!t]
  \centering
  \includegraphics[width=0.85\linewidth]{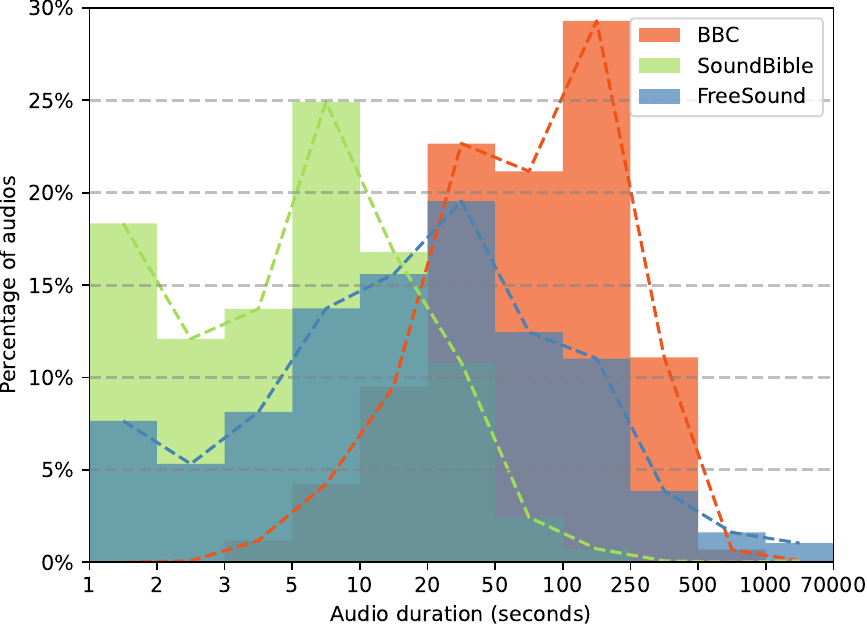}
  \caption{Distribution of audio duration from each data source in the WavCaps dataset. The audio clips from AudioSet are excluded since they are all 10 seconds long.}
  \label{fig:audio_dur}
\vspace{-15pt}
\end{figure}

\subsection{Dataset Analysis}
Table~\ref{tab:process_statistics} provides statistics for the raw data collected from four different sources before and after processing. It can be observed that more than half of the data samples from FreeSound were filtered out after processing, with the majority being removed by the high-frequency text filter. Since audio clips with a duration less than one second were excluded, the average duration of samples from FreeSound significantly increased. Conversely, only a small number of samples from BBC Sound Effects and SoundBible were removed. The dataset includes \num{28678} unique words and \num{19974} stemmed words. Notably, there has been a significant decrease in the average caption length.

Following \cite{lipping2019crowdsourcing}, we employed Jaccard similarity to quantify the overlap between the original raw descriptions and the captions augmented by ChatGPT. The Jaccard similarity is a metric used to gauge the similarity and diversity of sample sets. In our analysis, the Jaccard similarity is calculated as the proportion of the shared words to the total unique words in both the raw description and the ChatGPT-augmented caption. This can be formulated as:
\begin{equation}
    J(A, B) = \frac{|A \cap B|}{|A \cup B|}
    \label{eq:Jaccard}
\end{equation}
where $A$ is the set of words in the raw description and $B$ is the set of words in the ChatGPT-augmented caption. The normalized distribution of the Jaccard similarity scores, as presented in Figure \ref{fig:jaccard_similarities}, demonstrates a generally low level of lexical overlap across various sources. Such Jaccard similarity scores, coupled with the observed reduction in caption length, suggest that the ChatGPT-augmented captions have undergone considerable transformation from the original descriptions, often involving substantial word deletion. 

Fig.~\ref{fig:word_clouds} displays the top \num{100} word clouds in the WavCaps dataset and the entire harvested raw descriptions, respectively. The top word clouds in the WavCaps dataset contain meaningful sound-related words such as sounding objects and sound events, whereas the top word clouds in the raw descriptions are largely devoid of meaning. This also demonstrates that ChatGPT successfully extracted sound-related information and removed information unrelated to sound. Moreover, we utilized the Flesch-Kincaid grade level \cite{kincaid1975flesch} to evaluate the readability of both raw descriptions and ChatGPT-enhanced captions, where a lower Flesch-Kincaid grade level indicates the text is easier to read and comprehend. As depicted in Figure \ref{fig:flesch_kincaid}, the lower grade levels of the augmented captions across all the data sources reflect that ChatGPT has simplified the original raw descriptions. This simplification is aligned with our prompt design shown in Section \ref{ssec: data_char}, which aims to remove audio-irrelevant information and employ concise syntax.

\begin{figure*}[!htbp]
    \centering

    \begin{subfigure}[b]{0.3\textwidth}
        \includegraphics[width=\textwidth]{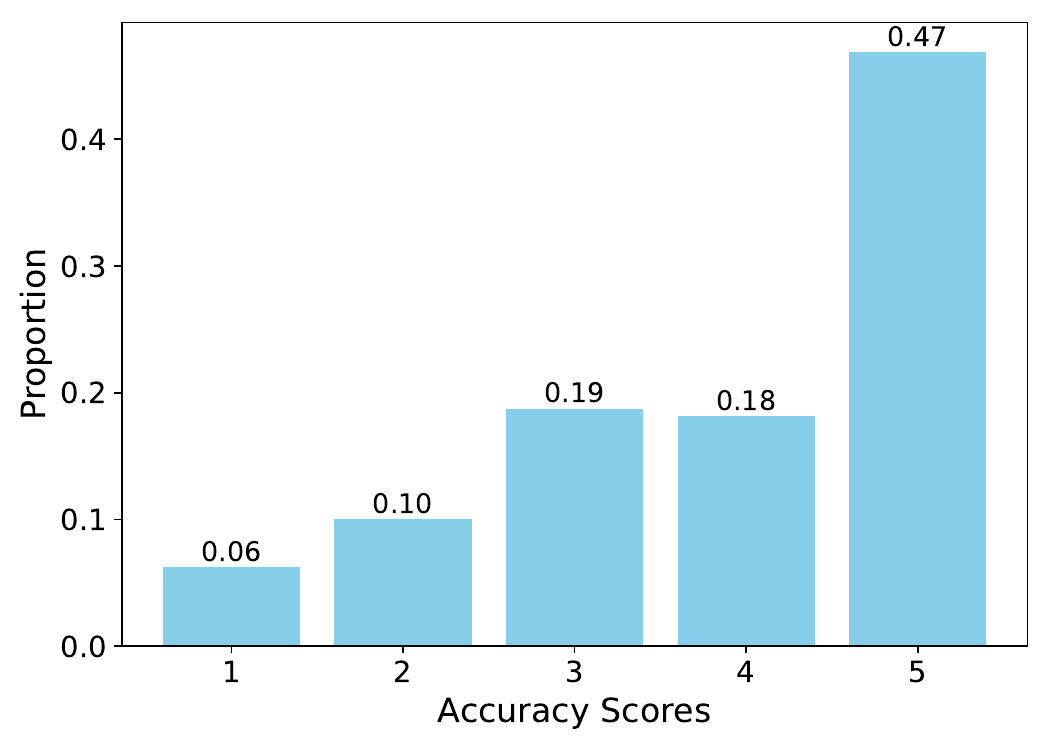}
        \caption{ }
        \label{fig:accuracy_scores}
    \end{subfigure}
    \begin{subfigure}[b]{0.35\textwidth}
        \includegraphics[width=\textwidth]{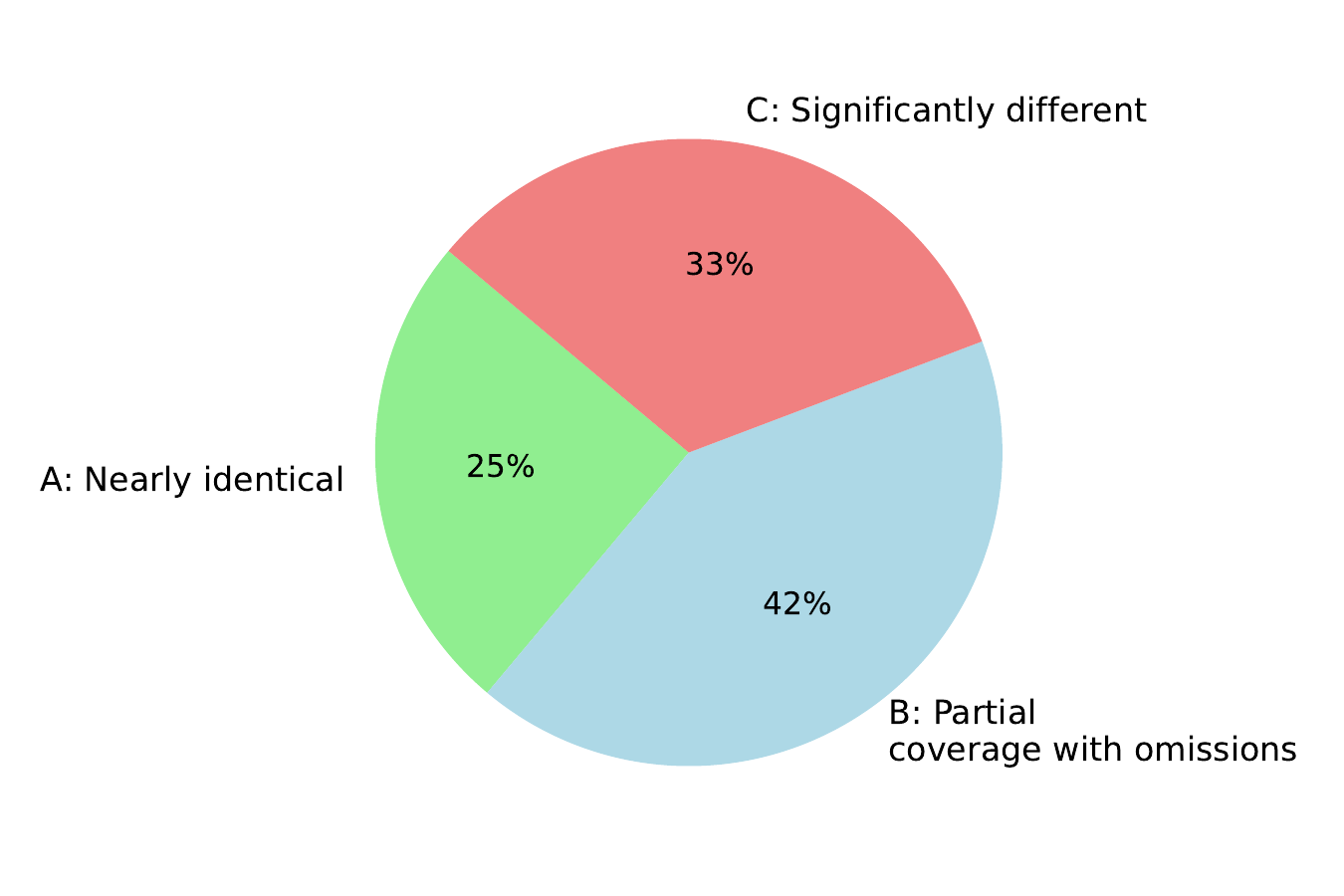}
        \caption{ }
        \label{fig:comparison_options}
    \end{subfigure}
    \begin{subfigure}[b]{0.3\textwidth}
        \includegraphics[width=\textwidth]{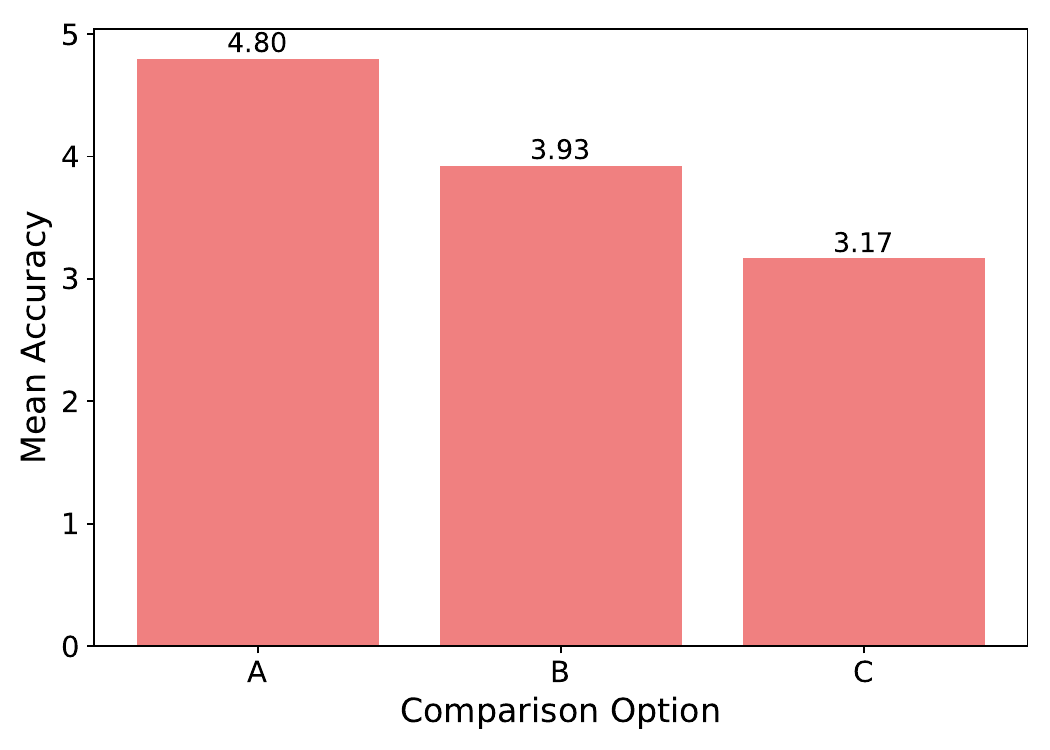}
        \caption{ }
        \label{fig:mean_accuracy_comparison}
    \end{subfigure}

    \caption{Visualization of human validation results. (a) The distribution of Mean Opinion Score (MOS) ratings for ChatGPT-augmented captions with 1 indicating no correspondence and 5 indicating an exact match to the audio content. (b) The evaluators' categorization of the similarity between ChatGPT-augmented captions and human-annotated captions: (A) nearly identical, (B) partial coverage with omissions, and (C) significantly different. (c) The mean accuracy scores of the captions corresponding to each of the three comparative categories between ChatGPT-augmented captions and human-annotated captions.}
    \label{fig:combined_figures}
\end{figure*}

Additionally, the WavCaps dataset includes \num{330609} unique captions. Among these, \num{311242} captions appear only once, showcasing a wide variety of expressions. However, there are \num{19367} captions that recur in the dataset, indicating instances where the exactly same caption (i.e., the same sequence of words) is used more than once. It is noteworthy that captions recurring with a frequency of more than \num{5} times tend to be shorter, having an average of \num{4.9} words, compared to the average length of all the captions, which have an average of \num{7.8} words. An informal analysis suggests that these shorter, frequently appearing captions typically depict common and single sound events. This trend suggests that, while there is considerable variation in the dataset, certain sound events tend to be described with a consistent, standardized vocabulary and captions.

Table~\ref{tab:dataset_statistics} provides a comparison of key statistics between WavCaps and other audio-language datasets. Human-labeled datasets are generally limited in size, but the captions are in high quality. SoundDescs \cite{koepke2022audioretrieval} is sourced from the BBC Sound Effects with no processing applied to the raw descriptions. LAION-Audio-630K has the highest number of audio clips. A majority of the audio clips in LAION-Audio-630K (around 420k) are sourced from FreeSound, resulting in a substantial overlap between their dataset and ours.  To preserve the diversity of clip durations, we did not exclude any long audio clips, leading to a longer total duration in our dataset compared to others. Fig.~\ref{fig:audio_dur} illustrates the distribution of audio durations in the WavCaps dataset. The AudioSet Strongly-labeled subset is not considered in Fig.~\ref{fig:audio_dur}, as all audio clips in AudioSet have a consistent duration of 10 seconds. Overall, in comparison to human-labelled audio captioning datasets, WavCaps is an order of magnitude larger and encompasses a greater diversity of content. In contrast to online-harvested datasets such as SoundDescs and LAION-Audio-630k, our method entails converting raw descriptions into captions and removing noisy or sound-unrelated descriptions. This approach ultimately leads to the development of the most extensive weakly-labeled audio captioning dataset that exists. 

We categorize WavCaps as a large-scale pre-training resource and have not divided it into subsets, in a similar way to the practice adopted in CC12M \cite{changpinyo2021cc12m}, or VideoCC 3M \cite{nagrani2022videocc3m}, for the reason that the dataset is collected from diverse sources and the captions are weakly-labelled. It is important to note that if the original meta descriptions do not capture a particular aspect of the audio events or even do not accurately describe the audio content, the resulting captions tend to reflect these limitations, as we only processed the text metadata, but not the audio data. In addition, high-quality audio captions may include spatial-temporal information about sound events, as observed in some human-annotated captions in datasets like Clotho and AudioCaps \cite{drossos2020clotho, kim2019audiocaps}. Such information can enhance the contextual understanding of the soundscapes. However, spatial-temporal details are generally absent in our harvested descriptions, particularly those sourced from three web sources. Given that spatial-temporal information is not a perquisite for effective audio captions, its omission aligns with our design principles. Consequently, our captions primarily lack spatial-temporal details, focusing instead on the coarse-grained descriptions of audio clips. Notably, the captions from AudioSet strongly-labelled subset might include temporal relationships, as these audio tags are provided to ChatGPT in sequence, based on their occurrence in time. In a future study, it would be interesting to further improve the granularity of the captions with the spatiotemporal information of the acoustic events.

\subsection{Human Validation}
To ensure the robustness of our methodology, we incorporated a human validation process for a subset of the WavCaps dataset. We randomly selected 160 samples from FreeSound, which are also included in the test set of Clotho. We recruited eight evaluators and engaged them in a two-stage process: initially, they were asked to listen to the audio samples and assessed the accuracy of the ChatGPT-augmented captions. The evaluators rated each caption on a Mean Opinion Score (MOS) scale from 1 to 5, where 1 means that the caption fails to correctly describe the audio clip at all, and 5 indicates a perfect description of the audio clip without errors. Subsequently, for the comparison phase, the evaluators were instructed to conduct a comparison of the ChatGPT-augmented captions against the human-annotated captions from the Clotho dataset. They were instructed to categorize the relationship between the two sets of captions from three options, (A) nearly identical in content, (B) ChatGPT-augmented captions partially covering the sound events as the Clotho annotations but omitting certain details or sound events, and (C) significantly different in content and detail. This comparative analysis was critical in identifying the relative accuracy and descriptive quality of the ChatGPT-augmented captions compared to human annotation standards. In the entire evaluation process, captions were labeled as ``Caption A'' and ``Caption B'' to ensure that evaluators remained unaware of each caption's origin, whether ChatGPT-augmented or human-annotated.

The human validation results are shown in Figure \ref{fig:combined_figures}. The mean accuracy score obtained for ChatGPT augmented captions was \num{3.89}, indicating a high level of precision in the ChatGPT-augmented captions. Notably, more than 40\% of the evaluated captions received a perfect score of \num{5}, indicating that the majority of the ChatGPT-augmented captions accurately describe the audio events. However, there were instances of lower scores, which underscores the rationale for labeling this dataset as weakly-labeled. As the captions are augmented from the raw descriptions, and if the raw descriptions do not cover all of the sound events, the augmented captions tend to miss such sound events too. In terms of the comparison options, the majority of evaluators selected option (B), suggesting that the ChatGPT-augmented captions usually missed some sound events or lacked the specificity found in human-annotated captions. This outcome is aligned with the initial intent of our prompt design, which intentionally discards specific details and generates more general captions. For option (C), which indicated a significant difference in content and detail, the mean accuracy score was \num{3.17}. If we regard the human-annotated captions as ground-truth, the mean accuracy score for option (C) is expected to be as low as possible. This discrepancy can partially be attributed to the annotation process used for Clotho dataset \cite{drossos2020clotho, lipping2019crowdsourcing}, where annotators had access solely to audio without contextual information, leading to possible interpretations of the same sound as different events. In conclusion, the human validation results affirm the overall quality of our dataset. The results highlight the efficacy of the ChatGPT-augmented captions that are generally accurate but could miss some details or sound events.

\section{Experiments}
\label{sec:exp}
To evaluate the impact of the proposed WavCaps dataset, we conducted experiments on several audio-language multimodal learning tasks, including audio-language retrieval, automated audio captioning, zero-shot audio classification, and text-to-sound generation. In this section, we provide a description of the AL tasks we considered, along with the corresponding experimental settings, results and analysis. For all the experiments excluding text-to-sound generation, audio clips are sampled with a \num{32}k Hz sampling rate, and we use 64-dimensional log mel-spectrograms extracted by a 1024-point Hanning window with a hop size of 320 samples as input audio features.

\begin{table*}[!t]
  \caption{Experimental results of audio-language retrieval on test sets of AudioCaps and Clotho. ``AC'' refers to ``AudioCaps'', ``LA'' refers to ``LAION-Audio-630K'', ``ZS'' refers to ``zero-shot", ``PT'' refers to ``pretraining'', and ``FT'' refers to ``fine-tuning''. Higher score means better performance.}
  \label{tab:retrieval_results}
  \centering
  \resizebox{\textwidth}{!}{
  \begin{tabular}{ccccc|ccc|ccc|ccc}
    \hline
    \multirow{3}{*}{\textbf{Model}} & \multirow{3}{*}{\textbf{Training Dataset}} & \multicolumn{6}{c}{\textbf{AudioCaps}} & \multicolumn{6}{c}{\textbf{Clotho}} \\
    \cline{3-14}  & & 
    \multicolumn{3}{c}{\textbf{Text-to-Audio}} & \multicolumn{3}{c}{\textbf{Audio-to-Text}} & \multicolumn{3}{c}{\textbf{Text-to-Audio}} & \multicolumn{3}{c}{\textbf{Audio-to-Text}} \\
    \cline{3-14}
    & & $\boldsymbol{R@1}$ & $\boldsymbol{R@5}$ & $\boldsymbol{R@10}$ & $\boldsymbol{R@1}$ & $\boldsymbol{R@5}$ & $\boldsymbol{R@10}$ & $\boldsymbol{R@1}$ & $\boldsymbol{R@5}$ & $\boldsymbol{R@10}$ & $\boldsymbol{R@1}$ & $\boldsymbol{R@5}$ & $\boldsymbol{R@10}$ \\
    \hline 
    MMT \cite{koepke2022audioretrieval}& AC or Clotho & 36.1 & 72.0 & 84.5 & 39.6 & 76.8 & 86.7 & 6.7 & 21.6 & 33.2 & 7.0 & 22.7 & 34.6 \\
    ML-ASE \cite{mei2022metric}& AC or Clotho & 33.9 & 69.7 & 82.6 & 39.4 & 72.0 & 83.9 & 14.4 & 36.6 & 49.9 & 16.2 & 37.6 & 50.2 \\
    TAP \cite{xin2023tap} & AC or Clotho & 36.1 & 72.0 & 85.2 & 41.3 & 75.5 & 86.1 & 16.2 & 39.2 & 50.8 & 17.6 & 39.6 & 51.4 \\
    CLAP-HTSAT \cite{deshmukh2022wavtext5k} & AC+Clotho+WavText5K & 34.6 & 70.2 & 82.0 & 41.9 & 73.1 & 84.6 & 16.7 & 41.1 & 54.1 & 20.0 & 44.9 & 58.7 \\
    LAION \cite{laionclap2023} & AC+Clotho & 36.7 & 70.9 & 83.2 & 45.3 & 78.0 & 87.7 & 12.0 & 31.6 & 43.9 & 15.7 & 36.9 & 51.3 \\
    LAION \cite{laionclap2023} & AC+Clotho+LA & 32.7 & 68.0 & 81.2 & 43.9 & 77.7 & 87.6 & 15.6 & 38.6 & 52.3 & 23.7 & 48.9 & 59.9 \\
    LAION  \cite{laionclap2023} & AC+Clotho+LA+AudioSet & 36.1 & 71.8 & 83.9 & 46.8 & 82.9 & 90.7 & 16.1 & 38.3 & 51.1 & 22.7 & 48.5 & 60.8 \\
    LAION (fusion) \cite{laionclap2023} & AC+Clotho+LA+AudioSet & 35.1 & 71.5 & 83.6 & 45.8 & 80.9 & 91.6 & 18.2 & 42.5 & 54.4 & 25.7 & 51.5 & 63.4 \\
    \hline
    HTSAT-BERT-ZS & LA & 14.4 & 39.4 & 53.9 & 15.3 & 40.1 & 56.2 & 13.8 & 35.5 & 47.4 & 15.8 & 39.7 & 51.8 \\
    HTSAT-BERT-PT & AC+Clotho+LA & 34.2 & 69.6 & 83.3 & 41.9 & 76.2 & 85.5 & 17.4 & 43.0 & 56.0 & 23.9 & 46.4 & 61.3 \\
    \hline
    CNN14-BERT & AC+Clotho & 30.3 & 66.1 & 79.7 & 36.7 & 70.6 & 83.0 & 15.6 & 39.7 & 52.6 & 19.9 & 41.2 & 55.9 \\
    CNN14-BERT-ZS & WavCaps & 25.4 & 58.2 & 73.2 & 32.4 & 63.6 & 77.1 & 17.5 & 41.7 & 54.9 & 21.7 & 45.7 & 57.6 \\
    CNN14-BERT-PT & AC+Clotho+WavCaps & 34.7 & 69.1 & 82.5 & 44.6 & 76.3 & 86.2 & 21.2 & 46.4 & 59.4 & 25.9 & 52.6 & 65.8 \\
    CNN14-BERT-FT & AC+Clotho+WavCaps & 35.1 & 70.0 & 82.1 & 45.7 & 76.1 & 87.7 & \textbf{21.5} & \textbf{47.9} & \textbf{61.9} & \textbf{27.1} & \textbf{52.7} & \textbf{66.3} \\
    HTSAT-BERT & AC+Clotho & 39.2 & 74.9 & 86.5 & 49.5 & 81.9 & 91.5 & 15.6 & 38.4 & 52.0 & 21.0 & 43.8 & 55.7 \\
    HTSAT-BERT-ZS & WavCaps & 28.6 & 61.1 & 75.8 & 40.2 & 69.4 & 80.3 & 16.5 & 38.8 & 50.9 & 20.0 & 43.3 & 56.6 \\
    HTSAT-BERT-PT & AC+Clotho+WavCaps & 39.7 & 74.5 & 86.1 & 51.7 & 82.3 & 90.6 & 19.5 & 45.2 & 58.2 & 23.4 & 50.9 & 63.4 \\
    HTSAT-BERT-FT & AC+Clotho+WavCaps & \textbf{42.2} & \textbf{76.5} & \textbf{87.1} & \textbf{54.6} & \textbf{85.2} & \textbf{92.4} & 19.7 & 45.7 & 59.4 & 26.9 & 52.6 & 64.9 \\
    \hline
  \end{tabular}
  }
\end{table*}

\subsection{Audio-Language Retrieval}
\label{ssec: al_retrieval}
Audio-language retrieval involves searching for an audio clip or a caption in a database based on a query from another modality. To perform this task, the model learns Acoustic Semantic Embeddings (ASE) \cite{mei2022metric} that map paired audio clips and captions closer in the embedding space, while keeping embeddings for non-paired audio clips and captions far apart. We evaluate audio-language retrieval under three training settings: zero-shot, pretraining, and fine-tuning.
 
\subsubsection{Models}
Following previous works \cite{laionclap2023, mei2022metric}, we build an ASE model based on a two-tower architecture, where an audio encoder is employed to encode audio representations while an language encoder is used to encode captions. Two types of audio encoders, a CNN14 from pretrained audio neural networks (PANNs) \cite{kong2020panns} and a Transformer network-HTSAT \cite{chen2022htsat} are considered, which are both pretrained on AudioSet with an audio tagging task. A pretrained BERT-base network \cite{devlin2019bert} is employed as the language encoder. A 2-layer multilayer perceptron with a ReLU activation in between is appended after these two encoders to project respective features into the shared embedding space. Cosine similarity is used to quantify the similarity between audio embeddings and language embeddings, and can be formulated as
\begin{equation}
    s_{ij} = \frac{f(a_i) \cdot g(t_j)}{||f(a_i)||_2||g(t_j)||_2}
    \label{eq:sim}
\end{equation}
where $f(\cdot)$ is the audio encoder and $g(\cdot)$ is the language decoder, $a_i$ is the  audio clip indexed with $i$ in a batch, $t_j$ is the caption indexed with $j$ in a batch, and $s_{ij}$ is the similarity score. The model is trained with a normalized temperature-scaled cross entropy loss (NT-Xent) \cite{chen2020ntxent} in a bi-directional manner, and can be formulated as:
\begin{equation}
\begin{split}
    \mathcal L  = -\frac{1}{2B}  \sum_{i=1}^{B} &\log \frac{\exp({s_{ii}/\tau})}{\sum_{j=1}^B\exp{({s_{ij}/\tau)}}} + \\
     & \log \frac{\exp({s_{ii}/\tau})}{\sum_{j=1}^B\exp{({s_{ji}/\tau)}}} 
    \label{eq:ntxent}
\end{split}
\end{equation}
where $B$ is the batch size, and $\tau$ is a temperature hyper-parameter. This training strategy is also known as contrastive language-audio pretraining~(CLAP) \cite{elizalde2022clap}.

\subsubsection{Experimental Setup}
We first train our two models on the merged training sets of AudioCaps and Clotho as baselines. For the zero-shot setting, we exclude all overlapping samples in AudioCaps and Clotho from the WavCaps dataset. We provide IDs of overlapping samples with the release of the dataset\footnote{\url{https://huggingface.co/datasets/cvssp/WavCaps/tree/main/json_files/blacklist}}. The baseline and zero-shot models are trained for \num{15} epochs with a batch size of \num{128} and a learning rate of \num{5e-05} using the Adam \cite{kingma2014adam} optimizer. For the supervised setting, we merge WavCaps, AudioCaps and Clotho together as a large training set (validation and test sets of AudioCaps and Clotho are not included). The models are trained for \num{40} epochs and other hyperparameters are the same as in zero-shot settings. For the fine-tuned setting, we aim to study the impact of pretraining on WavCaps dataset. We further fine-tune the models trained under pretraining setting on AudioCaps and Clotho for \num{20} epochs, respectively. The temperature hyperparameter $\tau$ is set to \num{0.07} for all settings. For the HTSAT audio encoder, all the audio clips are randomly cropped or padded to \num{10} seconds, because HTSAT requires fixed-sized inputs. For the CNN14 audio encoder that can receive variable length inputs, we set the maximum input duration as \num{30} seconds and audio clips longer than \num{30} seconds are randomly cropped. During training, audio clips with similar duration are grouped within a batch. Model checkpoints are selected based on their performance on validation sets after each epoch and the final model performance is evaluated on the test sets of AudioCaps and Clotho using recall at rank $k$ (R$@k$). For a query, R$@k$ is 1 if the positive item appears in the top $k$ retrieved items, otherwise 0. The final R$@k$ is averaged across the dataset.

\subsubsection{Results and Analysis}

Table \ref{tab:retrieval_results} presents the audio-language retrieval results on the AudioCaps and Clotho datasets, where top half of the table shows the results of existing methods and bottom half of the table shows our results. 
In the zero-shot setting, both of our models show strong ability for zero-shot retrieval. When compared with previous SOTA models that were trained only on respective datasets, both models outperform the previous SOTA models on the Clotho dataset and achieve comparable results on the AudioCaps dataset. This demonstrates that the models trained on WavCaps in zero-shot setting generalize well on both AudioCaps and Clotho datasets. In the pretraining setting, the performance is improved compared to baselines and zero-shot models, and also achieved SOTA results on both datasets. It can be observed that using more training data leads to greater improvement on audio-to-text metrics compared to text-to-audio metrics. With fine-tuning, both of our models further improve the performance, and outperform all existing methods by a significant margin on both audio-to-text retrieval ($16.7\%$ improvement on $R@1$ on AudioCaps and $5.4\%$ improvement on $R@1$ on Clotho) and text-to-audio retrieval ($15.0\%$ improvement on $R@1$ on AudioCaps and $18.1\%$ improvement on $R@1$ on Clotho). 

In order to demonstrate the effectiveness of our proposed WavCaps dataset, we primarily compare our results with models trained on the LAION-Audio-630K dataset \cite{laionclap2023}. For LAION's models, we observe a notable decline in performance on the AudioCaps dataset when the LAION-Audio-630K dataset is added to the training, while there is an improvement in results for the Clotho dataset. After incorporating AudioSet into the training, LAION's model demonstrates improved performance on AudioCaps, but remains close to their baseline. In our case, our models outperform LAION's models in most metrics for both datasets, despite utilizing less data. When incorporating WavCaps into the training, our models exhibit significant improvements on both datasets. It is important to highlight that LAION's training set, which includes AudioSet, is approximately six times larger than ours, totaling \num{2.63} million in size. Due to the differences in the architecture between our model and LAION's model, we trained our HTSAT-BERT model on the LAION-Audio-630K dataset under both zero-shot and pretraining settings. This approach ensures a fair comparison by maintaining consistency in all training parameters, with the only variation being the dataset used. The LAION-Audio-630K dataset we used contains \num{594404} audio clips, which is slightly smaller than the official dataset since we do not have access to some of the audio clips. The comparison between the official LAION-Audio-630K dataset and our version is shown in Appendix \ref{appendix}. It can be observed that the models trained on WavCaps greatly outperform those trained on LAION-Audio-630K under both training settings. Our experimental results reveal that the quality of the audio-language dataset significantly influences the model performance. By adopting our proposed three-stage processing pipeline to filter out noisy data and rewrite raw descriptions into captions, we achieve SOTA performance using a smaller quantity of higher-quality data.

Finally, our two models show varying performance levels on the AudioCaps and Clotho datasets. The CNN14-based model underperforms the HTSAT-based model on the AudioCaps dataset, yet it surpasses the HTSAT-based model on the Clotho dataset. A potential explanation for this could be the variable duration of audio clips in Clotho, which range from \num{15} to \num{30} seconds. Randomly cropping these clips into 10 seconds in the HTSAT model may result in information loss. As CNN14 is capable of handling variable duration, it outperforms HTSAT on the Clotho dataset. We believe that the performance of our models could be further improved by employing feature fusion methods, as seen in \cite{laionclap2023}, to process variable-duration audio clips. However, we plan to explore this in future work.

\begin{table*}[ht]
\caption{Automated audio captioning results on the test sets of AudioCaps and Clotho. A higher score means better performance. ``*" indicates that the validation set of Clotho is used to train the model.}
\label{table:caption_results}
\centering
\begin{tabular}[\linewidth]{ c | c | c c c c c c c} 
 \hline
 \textbf{Dataset} & \textbf{Model} &  \textbf{BLEU$_{1}$} & \textbf{BLEU$_{4}$} & \textbf{ROUGE$_l$} & \textbf{METEOR} & \textbf{CIDEr} & \textbf{SPICE} & \textbf{SPIDEr} \\ 
 \hline
 \multirow{12}{*}{Clotho} & Prefix \cite{kim2023prefix} & 56.0 & 16.0 & 37.8 & 17.0 & 39.2 & 11.8 & 25.5 \\
 & MAAC \cite{Ye2021peking} & 57.5 & 17.4 & 37.7 & 17.4 & 41.9 & 11.9 & 26.9 \\
 & Netease \cite{Han2021netease} & 58.3 & 17.7 & 38.8 & 17.9 & 45.6 & 12.8 & 29.2 \\
 & FeatureCut \cite{ye2022feature_cut} & \textbf{60.1} & 17.9 & 38.9 & 17.6 & 43.6 & 12.2 & 27.9 \\ 
  & CLIP-AAC* \cite{chen2022aac_contrastive} & 57.2 & 16.9 & 37.9 & 17.1 & 40.7 & 11.9 & 26.3 \\
 & Netease* \cite{Han2021netease} & 58.5 & \textbf{18.2} & \textbf{40.0} & 18.4 & 47.4 & \textbf{13.5} & 30.2 \\
 \cline{2-9}
 & CNN14-BART (zero-shot) & 29.9 & 7.2 & 29.3 & 12.0 & 24.8 & 8.7 & 16.7 \\
 & CNN14-BART (baseline)& 56.0 & 16.0 & 37.0 & 17.1 & 39.3 & 11.7 & 25.5 \\
 & CNN14-BART & \textbf{60.1} & 18.0 & \textbf{40.0} & \textbf{18.5} & \textbf{48.8} & 13.3 & \textbf{31.0} \\
 & HTSAT-BART (zero-shot) & 36.2 & 7.8 & 30.0 & 12.4 & 25.0 & 8.6 & 16.8 \\
 & HTSAT-BART (baseline) & 57.6 & 16.4 & 38.2 & 17.5 & 41.5 & 11.9 & 26.7 \\
 & HTSAT-BART & 58.5 & 16.8 & 38.3 & 18.4 & 46.2 & 13.3 & 29.7 \\ 
 \hline
 \multirow{11}{*}{AudioCaps} & ACT \cite{Mei2021ACT} & 64.7 & 25.2 & 46.8 & 22.2 & 67.9 & 16.0 & 42.0 \\
 & V-ACT \cite{liu2022vact} & 69.8 & 28.1 & 49.4 & 23.7 & 71.1 & 17.2 & 44.2 \\
 & Prefix \cite{kim2023prefix} & \textbf{71.3} & \textbf{30.9} & 50.3 & 24.0 & 73.3 & 17.7 & 45.5 \\
 & BART-tags \cite{Gontier2021ac_bart} & 69.9 & 26.6 & 49.3 & 24.1 & 75.3 & 17.6 & 46.5 \\
 & AL-MixGEN \cite{kim2022almixgen} & 70.0 & 28.9 & 50.2 & 24.2 & 76.9 & 18.1 & 47.5 \\
 \cline{2-9}
 & CNN14-BART (zero-shot) & 55.1 & 12.4 & 37.1 & 18.6 & 45.3 & 11.9 & 28.6 \\
 & CNN14-BART (baseline) & 67.0 & 26.1 & 48.3 & 23.1 & 72.1 & 16.9 & 44.5 \\
 & CNN14-BART & 69.3 & 27.2 & 49.9 & 24.7 & 75.6 & 17.9 & 46.8 \\
 & HTSAT-BART (zero-shot) & 51.3 & 11.0 & 37.8 & 20.4 & 39.3 & 13.8 & 26.7 \\
 & HTSAT-BART (baseline) & 67.5 & 27.2 & 48.3 & 23.7 & 71.1 & 17.7 & 44.4 \\
 & HTSAT-BART & 70.7 & 28.3 & \textbf{50.7} & \textbf{25.0} & \textbf{78.7} & \textbf{18.2} & \textbf{48.5} \\ 
 \hline
\end{tabular}
\end{table*}

\subsection{Automated Audio Captioning}
\label{sec:automated-audio-captioning}
Automated audio captioning is the task of generating a natural language sentence to describe the content of an audio clip, which mainly concerns environmental sounds and ignores possible voice content \cite{mei2022ac_review}. We continue to evaluate audio captioning on AudioCaps and Clotho datasets.

\subsubsection{Models}
Audio captioning is generally solved by an encoder-decoder model, where an encoder is leveraged to extract audio features and a decoder is employed to generate captions based on audio features extracted from the encoder. We build our model based on the baseline of Detection and Classification of Acoustic Scenes and Events (DCASE) 2022 challenge task 6 \cite{Gontier2021ac_bart}. Similar to the audio-language retrieval approach mentioned earlier, we explore two types of audio encoders, namely, a CNN14 and an HTSAT, for our audio captioning models. The language decoder is a pretrained language model, BART based network \cite{lewis2020bart}. BART is a sequence-to-sequence language model consisting of both Transformer encoder and decoder blocks, pretrained on large text corpora. The model is trained with a cross-entropy loss and can be formulated as:
\begin{equation}
  \label{eqn:ce_loss}
  \mathcal L_{\rm CE}(\theta)= - \frac{1}{T} \sum_{t=1}^T \log{p(y_t|y_{1:t-1}, x, \theta)}
\end{equation}
where $x$ is an input audio clip, $y_t$ is the $t$-th ground truth token in a sentence whose length is $T$, and $\theta$ are the parameters of the audio captioning model. 

\subsubsection{Experimental Setup}
We first explored a zero-shot training setting, where all overlapping samples from the Clotho and AudioCaps datasets are excluded for training. Then, we adopt a two-stage training paradigm similar to the fine-tuning setting in audio-language retrieval. The whole model is first pretrained on WavCaps, together with the training sets from Clotho and AudioCaps, using a learning rate of \num{5e-05} and a batch size of \num{48} for \num{15} epochs. The pretrained model is further fine-tuned on AudioCaps and Clotho for \num{20} epochs with a learning rate of \num{5e-06}, respectively. During the whole training process, we ensure that no data from validation or test sets of Clotho and AudioCaps are used for training. To assess the impact of pretraining using WavCaps dataset, we train the two models exclusively on AudioCaps and Clotho as baselines. The performance is evaluated using conventional metrics including BLEU$_n$ \cite{papineni2002bleu}, ROGUE$_l$ \cite{lin2004rouge}, METEOR \cite{banerjee2005meteor}, CIDEr \cite{vedantam2015cider}, SPICE \cite{anderson2016spice}, and SPIDEr \cite{liu2017spider}, where SPIDEr is generally employed as the main metric in the literature. 

\subsubsection{Results and Analysis}
Results are presented in Table~\ref{table:caption_results}. In the zero-shot training scenario, both models demonstrate robust zero-shot capabilities, affirming the suitability of our dataset for the audio captioning task. The fine-tuning results further demonstrate that our models surpass existing methods, achieving new SOTA performance on the Clotho and AudioCaps datasets. Notably, our model outperforms even those methods that incorporate the validation set into their training for the Clotho dataset. Compared to the baseline systems, pretraining on WavCaps leads to a significant improvement in the final performance on both datasets. These outcomes suggest that ChatGPT effectively transforms raw descriptions into caption-like sentences, thereby boosting audio captioning performance. In alignment with findings from audio-language retrieval above, the CNN14 audio encoder outperforms the HTSAT audio encoder on the Clotho dataset but exhibits inferior performance on the AudioCaps dataset. 

\subsection{Zero-Shot Audio Classification}
Audio classification aims at classifying the class of the sound presented in an audio clip. We carry out zero-shot audio classification on three popular audio event datasets to evaluate the  generalization and robustness of models trained on our WavCaps dataset.

\subsubsection{Models}
We formulate the audio classification task as an audio-language retrieval problem, following the approach as used in CLIP \cite{radford2021clip}. Initially, we encode all class labels as class embeddings using a text encoder without using any prompts. Each audio clip will be encoded by the audio encoder and then compared with the class embeddings to get a similarity score for each class. Those similarity scores will be normalized to get a final probability distribution over the classes. We use `HTSAT-BERT-PT' model described in Section~\ref{ssec: al_retrieval}. 

\subsubsection{Experimental Setup}
WavCaps and the training sets of AudioCaps and Clotho are merged as a training set, where overlapping samples co-occurred in the evaluation datasets are excluded. The training settings are same as those in audio-language retrieval. Three audio event datasets, ESC-50 \cite{piczak2015esc50}, UrbanSound8K \cite{salamon2014ub8k}, and VGGSound \cite{Chen2020vggsound} are employed. ESC-50 is an environmental sound classification dataset consisting of \num{2000} \num{5}-seconds audio clips annotated with \num{50} classes. UrbanSound8K contains \num{8732} audio clips less or equal to \num{4} seconds of urban sounds from 10 classes. VGGSound contains about \num{200}k audio clips for \num{310} classes sourced from YouTube videos. ESC-50 and UrbanSound8K are officially split into 5-folds and 10-folds for cross-validation. Therefore, we use all audio clips in ESC-50 and UrbanSound8K, and the test set of VGGSound for evaluation. Top-1 accuracy is used as the evaluation metric.

\begin{table}[!t]
\caption{Results of the top-1 accuracy on zero-shot audio classification.}
\centering
\resizebox{\linewidth}{!}{
\begin{tabular}{c|ccc}
\hline
\textbf{Model}  & \textbf{ESC-50} &\textbf{UrbanSound8K} & \textbf{VGGSound} \\
\hline
Supervised SOTA & 98.1 \cite{chen2022beats} & 90.0 \cite{gazneli2022strikes} & 75.4 \cite{laionclap2023} \\
\hline
Wav2CLIP \cite{wu2022wav2clip} & 41.4 & 40.4 & 10.0 \\
AudioCLIP \cite{guzhov2022audioclip} & 69.4 & 65.3 & - \\
CLAP \cite{elizalde2022clap} & 82.6 & 73.2 & - \\
BLAT \cite{xu2023blat}& 80.6 & 77.3 & 14.9 \\
LAION \cite{laionclap2023} & 91.0 & 77.0 & 29.1 (46.2) \\
\hline
Ours & \textbf{94.8} & \textbf{80.6} & \textbf{29.6} \\
\hline
\end{tabular}
}
\label{tab:zs_ac_results}
\end{table}

\subsubsection{Results and Analysis}
Table~\ref{tab:zs_ac_results} presents the results, with the top row showing the supervised SOTA performance on each dataset, while the remaining rows show zero-shot results. Compared to our model, LAION and BLAT \cite{xu2023blat} both use more data to train their models.
Our models achieved SOTA zero-shot results on all three datasets, significantly outperforming other models on ESC-50 and UrbanSound8K. In the case of VGGSound, Wu et al. \cite{laionclap2023} (LAION) reported a \num{46.2}\% accuracy, but their use of AudioSet for model training without excluding overlapping samples between AudioSet and VGGSound led to a data leakage issue. Overall, in comparison with LAION and BLAT, our approach yielded superior results using less data, which highlights the effectiveness of our proposed WavCaps dataset.

The zero-shot results we obtained were close to the supervised SOTA on ESC-50 and UrbanSound8K, but exhibited a considerable margin on the VGGSound dataset. ESC-50 and UrbanSound8K datasets are composed of 50 and 10 classes, respectively, while VGGSound dataset includes 310 classes. However, the level of granularity in VGGSound's classification scheme may be too high, which could lead to models trained on WavCaps struggling to generalize effectively across all 310 classes.

\begin{table}[!t]
\centering
\caption{Performance comparison on Text-based sound generation. AudioGen is marked with$^{\dag}$ because the pretrained model and the evaluation data of AudioGen are not open-sourced, which may lead to unreliable comparison.}
\label{tab: text-based-sound-generation-result}
\resizebox{\linewidth}{!}{
\begin{tabular}{lccccc}
\hline
\textbf{Model}                        & \textbf{Train Condition} & \textbf{FAD $\downarrow$}           & \textbf{IS $\uparrow$}            & \textbf{KL $\downarrow$}            & \textbf{FD $\downarrow$}            \\
\hline
    DiffSound~\cite{yang2022diffsound}                       & -              & 7.75 & 4.01 & 2.52 & 47.68         \\
    AudioGen-base$^{\dag}$~\cite{kreuk2022audiogen}                       & -              & 3.13 & - & 2.09 & -         \\
\hline
    \multirow{2}{*}{AudioLDM$_{\text{LAION}}$~\cite{liu2023audioldm}}                       & Audio              & 2.98          & \textbf{7.12} & 2.2           & 24.04         \\

    & Text               & 2.47          & 6.91          & 2.25          & 24.84         \\
\hline
\multirow{2}{*}{AudioLDM$_{\text{WavCaps}}$}       & Audio              & 2.85          & 5.42          & 2.31          & 24.89         \\
                            & Text               & \textbf{2.29} & 7.05          & 2.3           & 26.3          \\
\hline
\multirow{2}{*}{AudioLDM$_{\text{WavCaps-FT}}$} & Audio              & 2.98          & 5.79          & \textbf{2.17} & \textbf{21.9} \\
                            & Text               & 2.58          & 6.38          & 2.23          & 25.27      \\
\hline
\end{tabular}
}
\end{table}

\subsection{Text-based Sound Generation}

Text-based sound generation is a task that generates sound, including speech, music, and sound effect, with textual information~\cite{liu2023audioldm}. We follow previous studies~\cite{liu2023audioldm,kreuk2022audiogen,yang2022diffsound} and perform training and evaluation on the AudioCaps~\cite{kim2019audiocaps} dataset.


\subsubsection{Models} AudioLDM~\cite{liu2023audioldm} is a text-to-sound generation model that builds upon contrastive language-audio pretrained~(CLAP) encoders. When trained effectively, the CLAP encoders have the capability to capture the relationships between different modalities and ease the training challenges of audio generative models. The original AudioLDM, denoted by AudioLDM$_{\text{LAION}}$, adopts a CLAP model developed by~\cite{laionclap2023} with a dataset of around \num{2.6} million audio-text pairs. 
We re-implement AudioLDM with a CLAP model trained with our proposed WavCaps, denoted by AudioLDM$_{\text{WavCaps}}$. We further finetuned our CLAP model on AudioCaps training set for text-to-sound generation, denoted by AudioLDM$_{\text{WavCaps-FT}}$.
Specifically, we use the pretrained Variational Autoencoder~(VAE) and vocoder in the open-source implementation of AudioLDM\footnote{\url{https://github.com/haoheliu/AudioLDM}} and trained a new latent diffusion model~(LDM). The LDM in AudioLDM is trained with the re-weighted training objective~\cite{DDPM}, given by
\begin{align}
\label{trainingobjective}
L_{n}(\theta)&=\mathbb{E}_{\boldsymbol{z}_{0},\boldsymbol{\epsilon},n}\left \| \boldsymbol{\epsilon} - \boldsymbol{\epsilon}_{\theta}(\boldsymbol{z}_{n},n,\boldsymbol{E}^{x}) \right\|^2_{2},\\
q(\boldsymbol{z}_{n}|\boldsymbol{z}_{0})&=\mathcal  N(\boldsymbol{z}_{n};\sqrt{\bar{\alpha}_{n}}\boldsymbol{z}_{0},(1-\bar{\alpha}_{n})\boldsymbol{\epsilon}),
\end{align}
where $\epsilon \sim \mathcal{N}(0,\textit{\textbf{I}})$ denotes the standard Gaussian noise, $\bar{\alpha}_{n}$ is the noise schedule~\cite{liu2023audioldm}, $\boldsymbol{z}_{0}$ is the original VAE latent extracted from audio, and $\boldsymbol{z}_{n}$ is the output of the $n$-th forward diffusion step. 
To benchmark our models, we also included two state-of-the-art audio generation models, namely DiffSound~\cite{yang2022diffsound} and AudioGen~\cite{kreuk2022audiogen}.

\begin{table*}[!t]
  \caption{Ablation study results of zero-shot audio-language retrieval on the test sets of AudioCaps and Clotho.}
  \label{tab:ablation_results}
  \centering
  \resizebox{\textwidth}{!}{
  \begin{tabular}{cccc|ccc|ccc|ccc}
    \hline
    \multirow{3}{*}{\textbf{Settings}} & \multicolumn{6}{c}{\textbf{AudioCaps}} & \multicolumn{6}{c}{\textbf{Clotho}} \\
    \cline{2-13}& 
    \multicolumn{3}{c}{\textbf{Text-to-Audio}} & \multicolumn{3}{c}{\textbf{Audio-to-Text}} & \multicolumn{3}{c}{\textbf{Text-to-Audio}} & \multicolumn{3}{c}{\textbf{Audio-to-Text}} \\
    \cline{2-13}
    & $\boldsymbol{R@1}$ & $\boldsymbol{R@5}$ & $\boldsymbol{R@10}$ & $\boldsymbol{R@1}$ & $\boldsymbol{R@5}$ & $\boldsymbol{R@10}$ & $\boldsymbol{R@1}$ & $\boldsymbol{R@5}$ & $\boldsymbol{R@10}$ & $\boldsymbol{R@1}$ & $\boldsymbol{R@5}$ & $\boldsymbol{R@10}$ \\
    \hline 
    Before 1st step (FreeSound subset) & 12.9 & 37.0 & 51.8 & 13.8 & 38.2 & 55.1 & 12.5 & 36.2 & 46.9 & 15.4 & 36.8 & 48.3 \\
    After 1st step (FreeSound subset) & 15.6 & 43.2 & 59.6 & 19.4 & 46.8 & 62.8 & 14.3 & 36.8 & 49.4 & 16.5 & 39.0 & 52.3 \\
    After 3rd step (FreeSound subset) & 17.3 & 46.3 & 61.8 & 20.9 & 49.1 & 64.7 & 14.8 & 37.6 & 51.1 & 17.0 & 39.8 & 52.7 \\
    WavCaps & 28.6 & 61.1 & 75.8 & 40.2 & 69.4 & 80.3 & 16.5 & 38.8 & 50.9 & 20.0 & 43.3 & 56.6 \\
    \hline
  \end{tabular}
  }
\end{table*}

\begin{table*}[ht]
\caption{Ablation study results of zero-shot automated audio captioning on the test sets of AudioCaps and Clotho.}
\label{table:ablation_caption_results}
\centering
\begin{tabular}[\linewidth]{c | c | c c c c c c c} 
 \hline
 \textbf{Settings} & \textbf{Dataset} & \textbf{BLEU$_{1}$} & \textbf{BLEU$_{4}$} & \textbf{ROUGE$_l$} & \textbf{METEOR} & \textbf{CIDEr} & \textbf{SPICE} & \textbf{SPIDEr} \\ 
 \hline
 After 1st step (FreeSound subset) &  \multirow{2}{*}{Clotho} & 34.9 & 3.9 & 25.0 & 10.3 & 16.7 & 5.7 & 11.2 \\
 After 3rd step (FreeSound subset) & & 34.6 & 8.3 & 31.4 & 12.7 & 27.2 & 9.0 & 18.1 \\
 \hline
 After 1st step (FreeSound subset) &  \multirow{2}{*}{AudioCaps} & 
 30.5 & 3.0 & 24.6 & 10.7 & 21.2 & 5.8 & 13.5 \\
 After 3rd step (FreeSound subset) & & 34.0 & 6.8 & 30.0 & 12.1 & 36.2 & 7.7 & 22.0 \\
 \hline
\end{tabular}
\end{table*}

\subsubsection{Experimental Setup} 
The training and testing set split of AudioCaps are the same as the Audio Captioning experiments in Section~\ref{sec:automated-audio-captioning}. The LDM is optimized on the AudioCaps training set by an Adam optimizer with a learning rate of \num{3e-5}. We adopt a batch size of 8 and train LDM for a total of 400k steps. We evaluate the performance on the AudioCaps test set at intervals of 50k training steps and choose the outcome with the most optimal Frechet Audio Distance (FAD) as the final result for reporting purposes. The training data from AudioCaps are resampled to 16kHz before the model training. The setting on spectrogram calculation follows exactly the setting of~\cite{liu2023audioldm}. Regarding the AudioLDM-based models, we conduct experiments with two distinct modalities as training conditions: audio embedding and text embedding obtained from CLAP. When conditioned on audio, the model undergoes self-supervised training for audio generation, whereas conditioning on text involves supervised training using paired audio-text data.


In accordance with~\cite{liu2023audioldm} and their methodology, we assess the performance of our models using a range of metrics, including the FAD, Inception Score (IS), KL divergence (KL), and PANNS-based Frechet Distance (FD). The FAD and FD metrics evaluate the similarity between two audio data distributions, while IS measures the diversity of the generated audio data and its similarity to the target audio data distribution. Additionally, KL provides a sample-level measure of similarity between generated and target samples.

\subsubsection{Results and Analysis}

Table~\ref{tab: text-based-sound-generation-result} shows the evaluation result of AudioLDM paired with different CLAP models as well as the baseline methods. Even with a much smaller dataset size~(15\%) compared with AudioLDM$_{\text{LAION}}$, our AudioLDM$_{\text{WavCaps}}$ still achieves a comparable performance. AudioLDM$_{\text{WavCaps}}$ even performs better on FAD and the inception score when we use text as training embedding, indicating our dataset has better text labelling quality. When our CLAP model is finetuned on AudioCaps,  KL and FD by AudioLDM$_{\text{WavCaps-FT}}$ is improved significantly, which are even better than the KL and FD of AudioLDM$_{\text{LAION}}$.
Nevertheless, our model does not perform well on IS when conditioned on audio embedding. One possible explanation for this is that the scale of our training data is relatively small, which could restrict the model's ability to generalize. However, we can address this issue by augmenting the CLAP training data with label-to-caption augmented audioset data. Compared with the baseline methods, all our implementation of AudioLDM outperforms DiffSound and FAD of AudioGen by a large margin.

\subsection{Ablation Study}
\label{sec:ablation}
We carried out ablation studies to evaluate the effectiveness of each component in our processing pipeline. Recognizing that both step 2 (ChatGPT-based Transformation) and step 3 (Post-Processing) include caption refinement using ChatGPT, we merged these steps in our analysis. This approach enabled a direct comparison between the efficacy of the original raw descriptions and the ChatGPT-augmented captions. We specifically selected the FreeSound subset for our in-depth study, considering its notable complexity, and the fact that the first step of our pipeline primarily focuses on filtering data from FreeSound.

Table~\ref{tab:ablation_results} presents the experimental results of zero-shot audio language retrieval. The model used is `HTSAT-BERT' and the training settings are the same as those in the zero-shot setting in Section~\ref{ssec: al_retrieval}. Because we only used the FreeSound subset, the number of training samples is \num{561124} and \num{257040} before and after the 1st step of processing, respectively. It can be observed that all scores are improved after applying pre-filtering with less than half of the data. After the 3rd step, where the raw descriptions are augmented to captions by ChatGPT, slight improvements can be observed across all scores. When including data from other resources, all the scores are further improved, especially on AudioCaps.

The impact of caption refinement using ChatGPT in the audio-language retrieval task was relatively modest. Our primary motivation was to enhance raw descriptions into more comprehensive audio captions. To validate this approach, we conducted zero-shot automated audio captioning experiments. We trained the `CNN14-BART' model separately with the raw descriptions and the ChatGPT-augmented captions, using the FreeSound subset of the WavCaps dataset. Notably, the FreeSound subset, after the filtering of 1st processing step, comprised \num{257040} training samples. Table \ref{table:ablation_caption_results} shows the results of the ablation study for automated audio captioning. We can observe significant improvements after using the ChatGPT-augmented captions, where the SPIDEr score was improved by 6.9\% on the Clotho dataset and by 8.5\% on the AudioCaps dataset. This proves the effectiveness of the proposed ChatGPT-based caption refinement and the suitability of the WavCaps dataset for the audio captioning task.



\section{Conclusion}
\label{sec:conclu}
Data scarcity presents a significant challenge in audio-language multimodal learning research. In this study, we have introduced WavCaps, a large-scale weakly-labelled audio captioning dataset, created by collecting audio clips and their corresponding raw descriptions from the web. A three-stage processing pipeline is proposed to filter and transform crawled raw descriptions into captions using ChatGPT. Our evaluation of the WavCaps dataset on multiple audio-language multimodal learning tasks resulted in new state-of-the-art performance across all tasks. Our aspiration is that WavCaps can not only facilitate the progress of audio-language multimodal learning research but also showcase how ChatGPT-like LLMs can be utilized to enrich academic research.


\section*{Acknowledgments}
This work is supported partly by a Newton Institutional Links Award from the British Council, titled “Automated Captioning of Image and Audio for Visually and Hearing Impaired” (Grant number 623805725), and a grant EP/T019751/1 from the Engineering and Physical Sciences Research Council (EPSRC). For the purpose of open access, the authors have applied a Creative Commons Attribution (CC BY) licence to any Author Accepted Manuscript version arising. The authors wish to thank the associate editor and the reviewers for their helpful comments to further improve this work.



\appendix
\label{appendix}
\begin{table}[ht]
\centering
\caption{Comparison of the official LAION-Audio-630K and our downloaded version in terms of the number of audio clips and the number of captions per audio.}
\resizebox{\linewidth}{!}{
\begin{tabular}{lccc}
\toprule
Data Source & Official & Ours & Captions per audio \\
\midrule
FreeSound & 515581 & 501307 & 2 \\
BBC Sound Effects & 15973 & 15973 & 1 \\
Free To Use Sounds & 6370 & 6370 & 1 \\
Sonniss Game Effects & 5049 & 5561 & 1 \\
We Sound Effects & 488 & 488 & 1 \\
Paramount Motion Sound Effects & 4420 & 4932 & 1 \\
Audiostock & 10000 & 0 & 1 \\
Epidemic Sound & 75645 & 59773 & 2 \\
\bottomrule
\end{tabular}
}
\end{table}


\bibliographystyle{IEEEtran}
\bibliography{ref}

\vfill

\end{document}